\documentclass[sigconf,screen]{acmart}

\setcopyright{rightsretained}
\acmPrice{}
\acmDOI{10.1145/3445814.3446755}
\acmYear{2021}
\copyrightyear{2021}
\acmSubmissionID{asplos21main-p990-p}
\acmISBN{978-1-4503-8317-2/21/04}
\acmConference[ASPLOS '21]{Proceedings of the 26th ACM International Conference on Architectural Support for Programming Languages and Operating Systems}{April 19--23, 2021}{Virtual, USA}
\acmBooktitle{Proceedings of the 26th ACM International Conference on Architectural Support for Programming Languages and Operating Systems (ASPLOS '21), April 19--23, 2021, Virtual, USA}

\usepackage{tikz}
\usepackage{amsmath}

\usepackage{algorithm}
\usepackage{algpseudocode}
\usepackage{booktabs}
\usepackage{courier}
\usepackage{flushend}
\usepackage{forloop}
\usepackage{listings}
\usepackage{stackengine}
\usepackage{tikz}
\usepackage{xspace}
\usepackage[htt]{hyphenat}
\usepackage{enumitem}
\setlist[itemize]{leftmargin=*}

\newcommand*\circled[1]{\tikz[baseline=(char.base)]{\node[shape=circle,draw,inner sep=1pt] (char) {#1};}}

\newcommand{\term}[2]{\emph{#1}~(#2)}

\newcommand{\amorphos}{AmorphOS}
\newcommand{\cascade}{Cascade}

\newcommand{\quartus}{Quartus}
\newcommand{\system}{\xspace{\textsc{Synergy}}}

\newcommand{\verilog}{Verilog}
\newcommand{\vhdl}{VHDL}

\newcommand{\abi}{ABI}
\newcommand{\asic}{ASIC}

\newcommand{\cpu}{CPU}

\newcommand{\fpga}{FPGA}
\newcommand{\hdl}{HDL}

\newcommand{\ir}{IR}
\newcommand{\jit}{JIT}
\newcommand{\repl}{REPL}
\newcommand{\risc}{RISC}

\newcommand{\soc}{SoC}
\newcommand{\lut}{LUT}

\newcommand{\io}{IO}
\newcommand{\iop}{input/output}

\newcommand{\os}{OS}
\newcommand{\de}{DE10}
\newcommand{\mips}{MIPS}

\newcommand{\altera}{Altera}
\newcommand{\arm}{ARM}
\newcommand{\intel}{Intel}

\newcommand{\xilinx}{Xilinx}
\newcommand{\amazon}{Amazon}
\newcommand{\msft}{Microsoft}
\newcommand{\fone}{F1}
\newcommand{\stratix}{Stratix}
\newcommand{\avalon}{Avalon}

\definecolor{darksteelblue}{RGB}{73,131,178}
\lstdefinestyle{verilog} {language=Verilog, basicstyle=\fontfamily{courier}\ttfamily\small, keywordstyle=\bfseries\color{darksteelblue}, morekeywords={$fread,$feof}}
\lstdefinestyle{cpp} {language=C++, basicstyle=\fontfamily{courier}\ttfamily\small, keywordstyle=\bfseries\color{darksteelblue}}

\newcommand{\transformation}[3]{ %
        #1\parenVectorstack[l]{#2} & \Rightarrow & \Vectorstack[l]{#3}
}

\newcommand{\slowdown}{3-4}

\begin{document}

\date{}

\title{Compiler-Driven FPGA Virtualization with {\xspace{\textsc{SYNERGY}}}}

\author{Joshua Landgraf}
\affiliation{
  \institution{The University of Texas at Austin}
  \streetaddress{110 Inner Campus Drive}
  \city{Austin}
  \state{Texas}
  \postcode{78705}
  \country{USA}
}
\email{jland@cs.utexas.edu}

\author{Tiffany Yang}
\affiliation{
  \institution{The University of Texas at Austin}
  \streetaddress{110 Inner Campus Drive}
  \city{Austin}
  \state{Texas}
  \postcode{78705}
  \country{USA}
}
\email{tiffanyyang@utexas.edu}

\author{Will Lin}
\affiliation{
  \institution{The University of Texas at Austin}
  \streetaddress{110 Inner Campus Drive}
  \city{Austin}
  \state{Texas}
  \postcode{78705}
  \country{USA}
}
\email{wlsaidhi@utexas.edu}

\author{Christopher J. Rossbach}
\affiliation{
  \institution{The University of Texas at Austin \\ VMware Research Group \\ Katana Graph}
  \streetaddress{110 Inner Campus Drive}
  \city{Austin}
  \state{Texas}
  \postcode{78705}
  \country{USA}
}
\email{rossbach@cs.utexas.edu}

\author{Eric Schkufza}
\affiliation{
  \institution{Amazon}
  \city{Palo Alto}
  \state{California}
  \postcode{94303}
  \country{USA}
}
\email{eric.schkufza@gmail.com}

\begin{abstract}
\fpga{}s are increasingly common in modern applications, and
cloud providers now support on-demand FPGA acceleration in data
centers. Applications in data centers run on virtual infrastructure,
where consolidation, multi-tenancy, and workload migration enable
economies of scale that are fundamental to the provider's business. 
However, a general strategy for virtualizing \fpga{}s has yet to emerge.
While manufacturers struggle with hardware-based approaches,
we propose a compiler/runtime-based solution called \system{}.
We show a compiler transformation for \verilog{} programs that
produces code able to yield control to software at \emph{sub-clock-tick} granularity according to the semantics of the original program.  
\system{} uses this property to efficiently support 
core virtualization primitives: suspend and resume, program migration, 
and spatial/temporal multiplexing, on hardware which is available \emph{today}.
We use \system{} to virtualize \fpga{} workloads across a
cluster of \altera{} \soc{}s and \xilinx{} \fpga{}s on \amazon{}
\fone{}. The workloads require no modification, run within $\slowdown{}\times$ of
unvirtualized performance, and incur a modest increase in FPGA fabric utilization.

\end{abstract}



\maketitle

\section{Introduction}
\label{sec:intro}

Field-Programmable Gate Arrays (\fpga{}s) 
combine the functional efficiency of hardware with the programmability of
software.  \fpga{}s can exceed general-purpose \cpu{} performance by
orders of magnitude~\cite{catapult, configurable-cloud-acceleration}
and offer lower cost and time to market than \asic{}s. 
\fpga{}s
have become a compelling acceleration alternative for 
machine learning~\cite{brainwave, sharma2016dnnweaver, fpga-dnn,
fpga-opencl-cnn}, databases~\cite{fpga-db-accel, fpga-database-hashing,
fpga-kvs}, finance~\cite{cdo-pricing-fpga, hft-fpga}, graph
processing~\cite{fpgp, graphops},
communication~\cite{cheapbft,fpga-telco,fpga-openflow,consensus-in-a-box,catapult}, 
and image processing~\cite{neoface}.
In data centers, \fpga{}s serve diverse 
hardware needs with a single technology.
\amazon{} now provides \fone{} instances with large \fpga{}s attached~\cite{F1} and \msft{}
deploys \fpga{}s in new data center construction~\cite{olympus}.
 
Virtualization is fundamental to data centers. It 
decouples software from hardware, enabling economies of 
scale through consolidation. However, a standard technique for 
virtualizing \fpga{}s has yet to emerge. There are no widely agreed upon methods for 
supporting key primitives such as \term{workload
migration}{suspending and resuming a hardware program or
relocating it between \fpga{}s mid-execution} or
\term{multi-tenancy}{multiplexing multiple hardware
programs on a single \fpga{}}.
Better virtualization support is required for \fpga{}s to become a mainstream
accelerator technology.

Virtualizing \fpga{}s is difficult because they lack a well-defined interposable
\term{application binary interface}{ABI} and state capture primitives.
On CPUs, hardware registers are restricted to a small, static set and access to
data is abstracted through virtual memory, making it trivial to save and restore state.
In contrast, the state of an \fpga{} program is distributed throughout its
reprogrammable fabric in a \emph{program-} and \emph{hardware-dependent} fashion, 
making it inaccessible to the OS. 
Without knowing how programs are compiled for an \fpga{},
there is no way to share the \fpga{} with other 
programs or to relocate programs mid-execution.
\fpga{} vendors are pursuing hardware-based solutions to enable sharing
by partitioning the device into smaller, isolated fabrics.
However, lacking state capture primitives, this does not solve the fundamental problem 
and cannot support features like workload migration. 

We argue that the right place to support \fpga{} virtualization is in a
combined compiler/runtime environment.  Our system,
\system{}, combines a \term{just-in-time}{\jit{}} runtime for \verilog{},
canonical interfaces to OS-managed resources, 
and an OS-level protection layer to abstract and isolate shared resources.
The key insight behind \system{} is that a compiler can \emph{transparently} 
re-write \verilog{} code to compensate for the missing ABI and
explicitly expose application state to the OS.
The core technique in \system{} is a static analysis to transform the user's code into
a distributed-system-like \term{intermediate representation}{\ir{}} consisting
of monadic sub-programs which can be moved back and forth mid-execution between
a software interpreter and native \fpga{} execution.  This is possible because
the transformations produce code that can trap to
software at arbitrary execution points without violating the semantics of \verilog{}. 

\system{}'s first contribution is a set of compiler transformations to produce code that can be
interrupted at \emph{sub-clock-tick granularity}~(\S\ref{sec:prim}) according to the semantics of the original program.  This
allows \system{} to support a large class of \emph{unsynthesizable}
\verilog{}. Traditional \verilog{} uses
unsynthesizable language constructs for testing and debugging in a simulator:
\system{} uses these to expose interfaces to OS-managed resources and 
to start, stop, and save the
state of a program at any point in its execution.  This
allows \system{} to perform context switch and workload migration without hardware support or
modifications to \verilog{}. 

\system{}'s second contribution is a new technique
for \fpga{} multi-tenancy~(\S\ref{sec:ext}).  \system{}
introduces a hypervisor layer into the compiler's runtime which can combine the sub-program
representations from multiple applications
into a single hardware program, which is kept hidden from those instances.
This module is responsible for interleaving asynchronous data
and control requests between each of those instances and the \fpga{}.  In
contrast to hardware-based approaches, manipulating 
each instance's state is straightforward, as the hypervisor has access to every
instance's source and knows how it is mapped onto the device. 

\system{}'s final contribution is a compiler backend
targeting an OS-level protection layer
for process isolation, fair
scheduling, and cross-platform compatibility~(\S\ref{sec:impl}).
Recent OS-\fpga{} proposals 
harden vendor \emph{shells} and export interfaces for
the application to assist the OS with state capture for context switch~\cite{amorphos, optimus}.  
A major obstacle to using these systems is the requirement that the developer implement those state capture interfaces.
\system{} satisfies the state capture requirement transparently by using compiler
analysis to identify the set of variables
that comprise a program's state and emitting code 
to interact with state capture and quiescence interfaces.
For applications which natively support such interfaces, \system{} can use 
these to dramatically reduce overhead for context switch and migration. 

Our \system{} prototype extends the \cascade{}~\cite{cascade}
\jit{} compiler and composes it with the \amorphos{}~\cite{amorphos} \fpga{} OS. We measure \system{} in
real-world contexts that represent the heterogeneity of the data center.  We
show the ability to suspend and resume programs running on a cluster
of \altera{} \soc{}s and \xilinx{} \fpga{}s running on \amazon{}'s \fone{} cloud
instances, to transition applications between the two, and to temporally and
spatially multiplex both devices efficiently with strong \os-level
isolation guarantees.  This is done without exposing the
architectural differences between the platforms, or requiring extensions
to the \verilog{} language or modifications to the programs. 
We achieve performance within $\slowdown{}\times$ of unvirtualized code
with a reasonable fabric cost.

\section{Background}
\label{sec:bg}

\label{sec:bg_verilog}

\verilog{}~\cite{verilog} is one of two standard \hdl{}s used to program
\fpga{}s.  \vhdl{}~\cite{vhdl} is essentially isomorphic.  Verilog consists of
\emph{synthesizable} and \emph{unsynthesizable} constructs. Synthesizable
\verilog{} describes computation which can be lowered onto an \fpga{}.
Unsynthesizable \verilog{} includes tasks such as print statements, which
are more expressive and aid in debugging, but must be executed in software.

\begin{figure}[t]
\input{fig/verilog.v}
\caption{A simple \verilog{} module. \verilog{} supports a combination of sequential and concurrent semantics.}
\label{fig:bg_verilog}
\end{figure}

\verilog{} programs are declarative and organized hierarchically in units
called \emph{modules}.  An example \verilog{} module is shown in
Figure~\ref{fig:bg_verilog}. The interface to a module is defined in terms of
its \iop{} ports ({\tt clock}, {\tt res}). Its semantics are defined in
terms of arbitrary-width wires ({\tt x},{\tt y}) and registers ({\tt r}), logic
gates (e.g. {\tt \&}), primitive arithmetic (e.g. {\tt +}), and nested
sub-modules ({\tt sm}).  
The value of a wire is functionally determined by its inputs~(lines 5, 21),
whereas a register is updated at discrete intervals~(lines 6, 11,
13).  For brevity, our discussion ignores \verilog{}'s rules of type inference
({\tt reg} may be demoted to {\tt wire}).
\system{} \emph{does not}.

\verilog{} supports sequential and concurrent semantics.
Continuous assignments~(lines 5, 21) are scheduled
when the value of their right-hand-side changes. Procedural
blocks~(lines 9--19) are scheduled when their guard is satisfied (e.g. {\tt
clock} changes from 0 to 1).  The ordering of these events is
undefined, and their evaluation is non-atomic.  Any of the statements
in a {\tt fork/join} block may be evaluated in any order. Only a {\tt
begin/end} block is guaranteed to be evaluated sequentially. Procedural blocks
can contain two types of assignments to registers: blocking ({\tt =})
and non-blocking ({\tt <=}). Blocking assignments are executed immediately,
whereas non-blocking assignments must wait until all continuous
assignments or control blocks are finished. 

When used idiomatically, these semantics map directly onto hardware primtives:
wires \emph{appear} to change value instantly and registers \emph{appear} to
change value with the clock. However, unsynthesizable statements have no
analogue. The print statement on line 18 is non-deterministic, it can be
interleaved with any assignment in lines 10--14. So too is the first execution
of lines 12 and 14, which can be interleaved with the assignment on line 5.
While the assignment on line 11 is visible immediately, the assignment on line
13 is only performed after every block and assignment has been scheduled, thus
the value 3 only appears the second time line 10 is executed.

\subsection{Cascade}
\label{sec:bg_cascade}

\cascade{} is the first \jit{} compiler for \verilog{}. 
Using \cascade{}, \verilog{} is parsed one line at a time, added to the user's program, and its
side effects made visible immediately. This can include the results of
executing unsynthesizable \verilog{}.  While JIT compilation is orthogonal to
\system{}, \cascade{}'s runtime techniques are a fundamental building block. \cascade{} applies transformations to the user's program that
produce code which can trap into the \cascade{} runtime at the end of the
logical clock tick. These traps are used to handle unsynthesizable statements
in a way that is consistent with \verilog{}'s scheduling semantics, even during
hardware execution. \system{} improves upon this to trap into the runtime at
sub-clock-tick granularity according to the semantics of the original program and to enable context switch (\S\ref{sec:prim}).

\cascade{} uses the syntax of \verilog{} to manage programs at the module
granularity.  Its \ir{} expresses a distributed system of \verilog{}
\emph{sub-programs}, each corresponding to a single module in the user's
program.  A sub-program's state is represented by a data structure known as an
\emph{engine}.  Sub-programs start as low-performance, software-simulated
engines and are replaced over time by high-performance \fpga{}-resident
engines.  \cascade{} retains the flexibility to relocate engines by imposing a
constrained \abi{} on its \ir{}, mediated by messages sent over the runtime's
data/control plane.  Relevant to our discussion is a subset of that
\abi{}: {\tt get/set} and \texttt{evaluate/update} messages. The \texttt{get/set}
messages read and write an engine's inputs, outputs, and program variables.
The {\tt evaluate/update} messages request that an engine run until no more
continuous assigns or procedural blocks can be scheduled, and latch the result
of non-blocking assignments, respectively.

Unsynthesizable traps are placed on an ordered interrupt queue
and evaluated between clock ticks, when changes to engine state have
fixed-pointed and the program is in a consistent state. This limits support for
unsynthesizable \verilog{} to output-only. For example, print statements can
occur at any point in a program, but their side effects are only made visible
between clock-ticks.  There is no way to schedule an interrupt between the
statements in a {\tt begin/end} block, block on the result, and continue
execution. \system{} removes these limitations.

\subsection{AmorphOS}
\label{sec:bg_amorphos}

\amorphos{} is an \fpga{} runtime infrastructure which supports cross-program
protection and compatibility at very high degrees of multi-tenancy. \amorphos{}
allows hardware programs to scale dynamically in response to \fpga{} load and
availability and can transparently change mappings between user logic and
\fpga{} fabric to increase utilization by avoiding fragmentation.  
\amorphos{} extends processes with \emph{Morphlets}, an abstraction for
\fpga{}-based execution. \amorphos{} can spatially share an \fpga{} among
Morphlets from different protection domains and falls back to time-sharing when
space-sharing is infeasible.  \amorphos{} 
mediates \os{}-managed resources through a shell-like component called a hull,
which provides an isolation boundary and a compatibility layer.
This enables \amorphos{} to co-locate several Morphlets in a single
reconfigurable \emph{zone} to increase utilization without compromising security.
\amorphos{} leaves the problems of efficient context switch, over-subscription, and
support for multiple \fpga{}s mostly unsolved by relying on a
programmer-exposed quiescence interface and a programmer-populated compilation cache.

\amorphos{}'s
quiescence interface forces the programmer to write state-capture code
(\S\ref{sec:intro}), which requires explicitly identifying live state.  The interface is simple to
support for request-response style programs such as DNN inference
acceleration~\cite{sharma2016dnnweaver}, but difficult, say, for a \risc{} core
that can execute unbounded sequences of instructions.  This can subject an
\os{}-scheduler to arbitrary latency based on a program's implementation and
introduces the need for forced revocation mechanisms as a fallback.
Transparent state capture mechanisms which insulate the programmer from
low-level details of on-fabric state are not supported.

\section{Virtualization Primitives}
\label{sec:prim}

In this Section, we describe a sound transformation for \verilog{} that allows
a program to yield control at sub-clock-tick granularity. This transformation
allows \system{} to support the entire unsynthesizable \verilog{} standard from
hardware, including {\tt \$save} and {\tt \$restart}, the two primitives which
are necessary for supporting workload migration.  We frame this discussion with
a file \io{} case study. While file \io{} is not necessary for virtualization,
it provides a clear perspective from which to understand the transformation.
Moreover, supporting file \io{} in hardware enables a more expressive
programming environment in which applications have access to OS-managed
resources through canonical hardware-independent interfaces.  We leave a
discussion of other applications which can benefit from the ability to yield
control at the sub-clock-cycle granularity (say, step-through debuggers) to
future work.

\begin{figure}[t]
\input{fig/file_io.v}
  \caption{Motivating example. A \verilog{} program that uses unsynthesizable \io{} to sum the values in a large file.}
\label{fig:prim_file_io}
\end{figure}

\subsection{Motivating Example: File I/O}

Consider the program shown in Figure~\ref{fig:prim_file_io}, which uses
unsynthesizable \io{} tasks to sum the values contained in a large file. The
program opens the file (line 4) and on every clock tick, attempts to read a
32-bit value (line 9). When the program reaches the end-of-file, it prints the
running sum and returns control to the host (lines 10-12).  Otherwise, it adds
the value to the running sum and continues (line 14). While this program is
simple, its structure is typical of applications that perform streaming
computation over large data-sets. 

The key obstacle to supporting this program is that the \io{} tasks
introduce data-dependencies within a single clock-tick. The end-of-file
check on line~10 depends on the result of the read operation on line~9, as does
the assignment on line~14. Because the semantics of these operations involve an
interaction with the file system, we must not only pause the execution of the
program mid-cycle while control is transferred to the host, but also block for
an arbitrary amount of time until the host produces a result. Our solution is
to transform the program into a state machine which implements a co-routine
style semantics. While a programmer could adopt this idiom, the changes would harm both readability and
maintainability. 

\begin{figure}[t]
  \small
  \begin{displaymath}
    \begin{array}{rcl}
      \mathcal{S}\big({\tt fork} \ s_1 \dots s_n \ {\tt join}\big) & \Rightarrow & {\tt begin} \ s_1 \dots s_n \ {\tt end}
      \\[.5em]
      \mathcal{S}\big({\tt begin} \ s_1 \dots s_n \ {\tt end}\big) & \Rightarrow & \mathcal{S}(s_1) \dots \mathcal{S}(s_n)
      \\[.5em]
      \transformation
        {\mathcal{S}}
        {{{\tt always \ @(}\epsilon_1{\tt)} \ s_1} {\dots} {{\tt always \ @(}\epsilon_n{\tt)} \ s_n}} 
            {{{\tt always \ @(}\epsilon_1 { \ {\tt or}} \dots { \ {\tt or}} \ \epsilon_n {\tt)}}   { \ \ {\tt if \ (}\mathcal{G}\big(\epsilon_1\big){\tt)} \ {\tt begin} \ \mathcal{S}(s_1) \ {\tt end}}  { \ \ \dots} { \ \ {\tt if \ (}\mathcal{G}\big(\epsilon_n\big){\tt)} \ {\tt begin} \ \mathcal{S}(s_n) \ {\tt end}}}
      \\[2.5em]
      \mathcal{S}\big(s\big) & \Rightarrow & s
      \\[.5em]
      \mathcal{G}\big({\tt posedge \ x}\big) & \Rightarrow & {\tt \_\_pos\_x}
      \\[.5em]
      \mathcal{G}\big({\tt negedge \ x}\big) & \Rightarrow & {\tt \_\_neg\_x}
      \\[.5em]
      \mathcal{G}\big({\tt x}\big) & \Rightarrow & {\tt \_\_any\_x}
    \end{array}
  \end{displaymath}
  \caption{Transformations used to establish the invariant that procedural logic appears in a single control statement.}
\label{fig:prim_sched}
\end{figure}

\subsection{Scheduling Transformations}

\system{} uses the transformations sketched in Figure~\ref{fig:prim_sched} to
establish the invariant that all procedural logic appears in a single control
statement.
Any {\tt fork/join} block may be replaced by an equivalent {\tt begin/end}
block, as the sequential semantics of the latter are a valid scheduling of the
former.  Also, any nested set of {\tt begin/end} blocks may be flattened into a
single block as there are no scheduling constraints implied by nested blocks.
Next, we combine every procedural control statement in the program into a
single statement called \emph{the core}. The core is guarded by the union of
the events that guard each individual statement. This is sound, as \verilog{}
only allows disjunctive guards. Next, we set the body of the core to a new {\tt
begin/end} block containing the conjunction of the bodies of each individual
block. This is sound as well, as sequential execution is a valid scheduling of
active procedural control statements.  Finally, we guard each conjunct with a
name-mangled version of its original guard (e.g.  {\tt \_\_pos\_x}, details
below) as all of the conjuncts would otherwise be executed when the core is
triggered. We note that these transformations are sound even for programs with
multiple clock domains.

\begin{figure}[t]
  \small
  \begin{displaymath}
    \begin{array}{rcl}
      \mathcal{\delta}\big({\tt x}\big) & \Rightarrow &
      \Vectorstack[l]{{\tt reg \ \_\_px;} {\tt always \ @(posedge \ \_\_clk)} {\tt \ \ \_\_px \ <= \ x;} }
      \\[-.5em]
      \mathcal{D}\big({\tt posedge \ x}\big) & \Rightarrow &
      \Vectorstack[l]{\mathcal{\delta}\big(x\big) {\tt wire \ \_\_pos\_x \ = \ !\_\_px \ \& \ x;}} \\
      \\[-.5em]
      \mathcal{D}\big({\tt negedge \ x}\big) & \Rightarrow &
      \Vectorstack[l]{\mathcal{\delta}\big(x\big) {\tt wire \ \_\_neg\_x \ = \ \_\_px \ \& \ !x;}} \\
      \\[-.5em]
      \mathcal{D}\big({\tt x}\big) & \Rightarrow &
      \Vectorstack[l]{\mathcal{\delta}\big(x\big) {\tt wire \ \_\_any\_x \ = \ \_\_px \ != \ x;}} \\
      \\[-.5em]
      \mathcal{C}\big({\tt always \ @(}\mathcal{E}{\tt )} \ s\big) & \Rightarrow &
            \Vectorstack[l]{{\forall_{\epsilon \in \mathcal{E}} \ \mathcal{D}(\epsilon)} {\tt reg[31:0] \ \_\_state;} {\tt reg[31:0] \ \_\_task;} {\tt always @(posedge \ \_\_clk)} { \ \ Lower(\mathcal{M}(s))}} \\
    \end{array}
  \end{displaymath}
  \caption{Transformations used to convert the core control statement into a form compatible with the \system{} \abi{}.}
\label{fig:prim_control}
\end{figure}

\subsection{Control Transformations}

The transformations in Figure~\ref{fig:prim_control}
modify the control structure of the core so that it is compatible with
the \cascade{} \abi{}.  Recall that the \cascade{} \abi{} requires that all of the inputs to an \ir{} sub-program
\emph{including clocks} will be presented as values contained in {\tt set}
messages which may be separated by many native clock cycles on the target
device. Thus we declare state to hold the previous values of
variables that appear in the core's guard, and wires that capture their
semantics in the original program (e.g.  {\tt \_\_pos\_x} is true whenever a
{\tt set} message last changed {\tt x} from false to true. We also declare new
variables ({\tt \_\_state} and {\tt \_\_task}) to track the control state of
the core, and whether a system task requires the attention of the runtime.
Finally, we replace the core's guard by a posedge trigger for the native clock
on the target device ({\tt \_\_clk}).

\begin{figure}[t]
\input{fig/file_io_res.v}
  \caption{The motivating example after modification to yield control to the runtime at the sub-clock-tick granularity.}
\label{fig:prim_file_io_res}
\end{figure}

\subsection{State Machine Transformations}

The body of the core is lowered onto a state machine with
the following semantics.
States consist of as many synthesizable statements as possible and are
terminated either by unsynthesizable tasks or the guard of an {\tt if} or {\tt
case} statement. A new state is created for each branch of a conditional
statement, and an SSA-style phi state is used to rejoin control flow.

A compiler has flexibility in how it chooses to lower the resulting state
machine onto \verilog{} text. Figure~\ref{fig:prim_file_io_res} shows one
possible implementation. Each state is materialized as an \texttt{if} statement
that performs the logic associated with the state, takes a transition, and sets
the {\tt \_\_task} register if the state ended in an unsynthesizable statement.
Control enters the first state when the variable associated with the original
guard ({\tt \_\_pos\_clock}) evaluates to true, and continues via the
fall-through semantics of \verilog{} until a task is triggered. When this
happens, a runtime which is compatible with the \cascade{} \abi{} can take
control, place its results (if any) in the appropriate hardware location, and
yields back to the target device by asserting the {\tt \_\_cont} signal. When
control enters the final state, the program asserts the {\tt \_\_done} signal,
indicating that there is no further work to do be done. Collectively, these
steps represent the compute portion of the {\tt evaluate} and {\tt update}
requests required by the \abi{}.

\subsection{Workload Migration}

With these transformations, support for the {\tt \$save} and {\tt \$restart}
system tasks is straightforward. Both can be materialized as traps triggered by
the value of {\tt \_\_task} in a runtime compatible with the \cascade{} \abi{}.
The former prompts the runtime to save the state of the program through a
series of {\tt get} requests, and the latter prompts a sequence of {\tt set}
requests. Either statement can be triggered in the course of normal program
execution, or via an eval statement.
Once a program's state is read out, it can be suspended, migrated to another
physical machine if necessary, and resumed.

\section{Hypervisor Design}
\label{sec:ext}

In this section we describe \system{}'s support for the two primary forms of
hardware multiplexing: \term{spatial}{where two programs are run simultaneously
on the same fabric} and \term{temporal}{where two programs share resources
using time-slice scheduling}.  \system{} provides an indirection layer that
allows multiple runtime instances 
to share a compiler at the hypervisor layer. 

\subsection{Program Coalescing}
\label{sec:ext_multi}

\begin{figure}[t]
\centering
\includegraphics[scale=1.0]{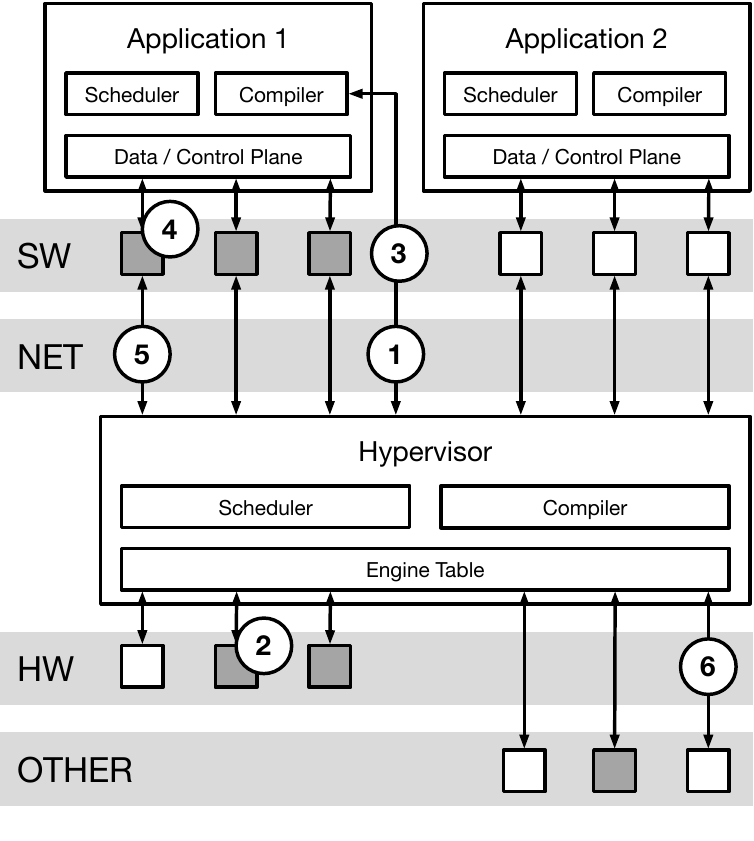}
	\caption{The \system{} virtualization layer. Sub-programs from multiple applications are combined on one target.}
\label{fig:ext_hyper}
\end{figure}

Figure~\ref{fig:ext_hyper} shows a sketch of \system{} during an 
execution in which two applications share a single hardware
fabric. In addition to the scheduler and data/control plane introduced in
\S~\ref{sec:bg}, we have called out the compilers associated with
both the runtime instance running those applications, and the \system{} hypervisor. 
These compilers are responsible for lowering a sub-program onto a
target-specific engine that satisfies \cascade{}'s distributed-system \abi{}. 

The compiler in the runtime instance
connects to the hypervisor~\circled{1}, which runs
on a known port. It sends the source code for a sub-program over the
connection, where it is passed to the native hardware compiler in the hypervisor,
which produces a target-specific implementation  of an engine and places it on the
\fpga{} fabric~\circled{2}.  The hypervisor 
responds with a unique identifier representing the engine~\circled{3} and
the runtime's compiler creates an engine
which remains permanently in software and is configured with the unique
identifier~\circled{4}. The resulting engine interacts
with the runtime as usual. However, its implementation of the
\cascade{} \abi{} is simply to forward requests across the network to 
the hypervisor~\circled{5} and block further execution until a reply is obtained.

The key idea that makes this possible is that the compiler in
the hypervisor has access to the source code for every sub-program in every
connected instance. This allows the compiler to support multitenancy
by combining the source code for each sub-program into a single monolithic
program. Whenever the text of any sub-program changes, the combined
program is recompiled to support the new logic. Whenever an
application finishes executing, all of its sub-programs are flagged for removal on the next recompilation. 
The implementation of this combined program is straightforward.
The text of the sub-programs is placed in modules named after their unique
hypervisor identifier. The combined program concatenates these modules together
and routes \abi{} requests to the appropriate module based on their identifier.
By isolating both sub-program code and communication, the FPGA fabric can be shared securely.

The overhead of the \system{} hypervisor depends primarily on the application.
While regular communication can become a bottleneck,
optimizations~\cite{cascade} can reduce the \abi{} requests between the runtime
and an engine to a tolerable level.  For batch-style applications, fewer than
one \abi{} request per second is required, and we are able to achieve
near-native performance even for programs separated from the hypervisor by a
network connection.  In contrast, applications that invoke frequent ABI calls
(e.g. for file I/O) will have overheads that scale with the frequency of
interaction. 
While our discussion presents a hypervisor which
compiles all of its sub-programs to \fpga{} fabric, this is not fundamental. 
The virtualization layer nests, and it is both possible and
performant for a hypervisor to delegate the compilation of a sub-program to a
second hypervisor~\circled{6}, say if the device is full.

\subsection{Scheduling State-Safe Compilation}
\label{sec:ext_int}

The \system{} hypervisor schedules \abi{} requests sequentially to avoid
resource contention. The one exception is compilation, which can
take a very long time to complete. If compilation were serialized
between \abi{} requests, it could render applications
non-interactive. But scheduling compilation asynchronously leads to a key
implementation challenge: 
changing the
text of one instance's sub-programs requires that the entire
\fpga{} be reprogrammed, a process which would destroy all connected instances'
state. The solution is to schedule these
destructive events when all connected instances are between
logical clock-ticks and have saved their state.

\begin{figure}[t]
\centering
\includegraphics[scale=1.0]{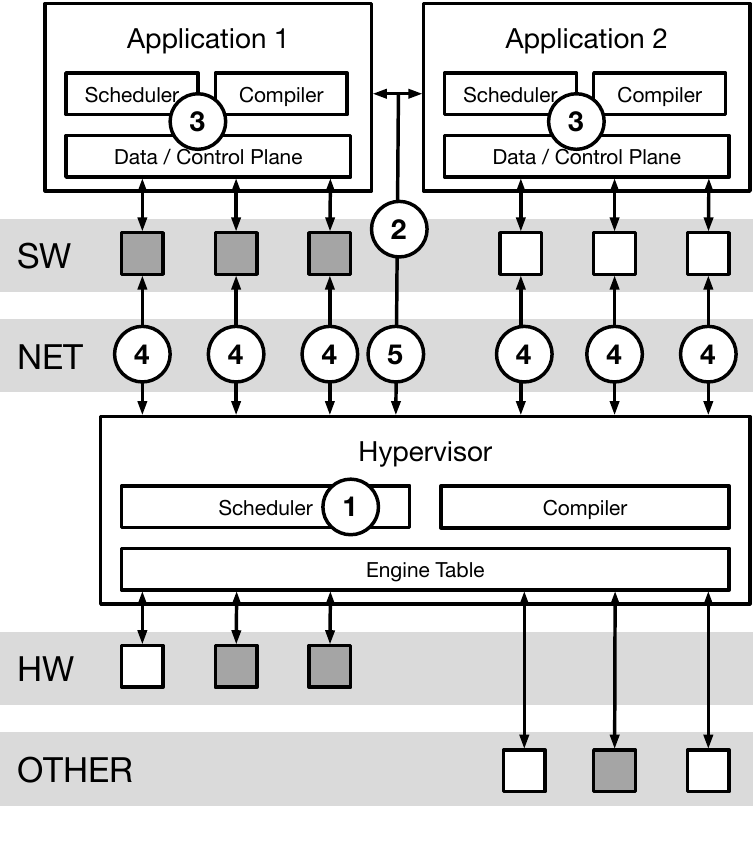}
\caption{The handshake protocol used to establish state-safe interrupts in the \system{}'s scheduler.
}
\label{fig:ext_handshake}
\end{figure}

Figure~\ref{fig:ext_handshake} shows the handshake protocol used to establish
these invariants. Compilation requests are scheduled
asynchronously~\circled{1}, and run until they would do something destructive.
The hypervisor then sends a request to every connected runtime
instance~\circled{2} to schedule an interrupt between their logical clock-ticks
when they are in a consistent state~\circled{3}.  The interrupt causes the
instances to send {\tt get} requests to \system{}~\circled{4} to save their
program state. When they have finished, the instances send a reply indicating
it is safe to reprogram the device~\circled{5} and block until they receive an
acknowledgement.  Compilation proceeds after the final reply.  The device is
reprogrammed and the handshake finishes in the opposite fashion. The hypervisor
informs the instances it is finished, they send {\tt set} requests to restore
their state on the target device and control proceeds as normal. 

\subsection{Multitenancy}

Collectively, these techniques suffice to enable multitenancy. Spatial multiplexing is accomplished by
combining the sub-programs from each connected runtime into a single
monolithic program on the target device. Temporal multiplexing is accomplished
by serializing \abi{} requests that involve an \io{} resource (say,
a connection to an in-memory dataset) which is in use by another sub-program.
Sharing preserves tenant protection boundaries using \amorphos{}, which provides support
for isolating sub-programs sharing the \fpga{} fabric (\S\ref{sec:bg_amorphos}).

\section{Implementation}
\label{sec:impl}

Our implementation of \system{} comprises the hypervisor described in
\S\ref{sec:ext}, compilation passes which enable sub-clock-tick granularity
support for the unsynthesizable primitives described in \S\ref{sec:prim}, and
both \intel{} and \amorphos{} backends. 

\subsection{\intel{} Backends}

Our implementation of \system{} extends \cascade{}'s support for the \de{} Nano
\soc{} to the full family of \intel{} devices that feature reprogrammable
fabric and an \arm{} core. This describes a range of targets, including the
high-performance \stratix{} 10. The core feature
these targets share is that they support \intel{}'s \avalon{}
interface for memory-mapped \io{}. This allows us to lower the
transformations described in \S\ref{sec:prim} onto a \verilog{} module
that converts reads and writes on the \avalon{} memory-mapped slave interface
into \abi{} requests. 

Adding support for a new \intel{} backend amounts to compiling this
module in a hardware context which contains an \avalon{} memory-mapped
master whose control registers are {\tt mmap}'ed into the same process space as
the runtime or hypervisor.  Compiling the logic for these interfaces can be
expensive, so \system{} augments the \intel{} family of backends with a
compilation cache similar to the one used by \amorphos{}. This allows \system{}
to transition gracefully from using \cascade{}'s \jit{} interface for
iterative development to using fast database lookup for mature virtualized
applications that require rapid transitions to hardware execution.
Unlike the \amorphos{} backend described below, 
our \de{} backend does not yet support the \amorphos{} protection layer.

\subsection{\amorphos{} Backends}
\label{sec:impl_f1}

\system{} uses a similar strategy for supporting multiple \amorphos{} backends.
We lower the transformations described in \S\ref{sec:prim} onto a \verilog{}
module implementing the \amorphos{} CntrlReg interface.  The module runs as a
Morphlet inside the \amorphos{} hull, which provides cross-domain protection
and thus preserves tenant isolation boundaries. It also enables \system{} to
take advantage of the large degree of multitenancy \amorphos{} offers.
The \system{} hypervisor
communicates with the Morphlet via a library from \amorphos{}.  This makes
adding support for a new \amorphos{} backend as simple as bringing \amorphos{}
up on that target.

A key difference between the \de{} and \fone{} is the size and speed of the
reprogrammable fabric they provide. Each \fone{} \fpga{} has $10\times$
more \lut{}s and operates $5\times$ faster than a \de{}.  This enables
\system{} to accelerate larger applications, but also makes achieving timing
closure challenging. \system{} adopts two solutions.  The first is to pipeline access
to program variables which are modified by {\tt get/set} requests.  For writes,
\system{} adds buffer registers between the \amorphos{} hull and the
variables.  For reads, \system{} builds a tree with the program's
variables at the leaves and the hull at the trunk. By adding
buffer registers at certain branches, this logic is removed from the critical
timing path.  The second solution is to iteratively reduce the target device
frequency until the design does meets timing.  This is automated by \system{}'s
build scripts, which can also preserve synthesis, placement, and routing data
to help offset the cost of performing multiple compiles.

\begin{figure}[t]
\input{fig/quiescence.v}
        \caption{The {\tt \$yield} task enables \system{}'s quiescence
        interface.  Volatile variables must be managed by the user.}
\label{fig:impl_q}
\end{figure}

\subsection{Quiescence Interface}

\amorphos{} provides a quiescence interface that notifies applications when
they will lose access to the FPGA (e.g. during reconfiguration),
allowing them to quiesce and back up their state accordingly.
\system{} supports this interface by handling the implementation of
execution control and state management for developers.
By default, all program variables are considered {\tt non\_volatile},
and will be saved and restored automatically.

For applications that implement quiescence, \system{} introduces
an optional, non-standard {\tt \$yield} task, shown in Figure~\ref{fig:impl_q}.
Developers can assert {\tt \$yield} to signal that the program has entered an
application-specific consistent state. When present, \system{} will only
perform state-safe compilations at the end of a logical clock tick in which
{\tt \$yield} was asserted.
The use of {\tt \$yield} causes stateful program variables to be considered
\emph{volatile} by default. Volatile variables are ignored by state-safe
compilations, making it is the user's responsibility to restore or reset their
values at the beginning of each logical clock tick following an invocation
of {\tt \$yield}.  Users may override this behavior by annotating a variable
as {\tt non\_volatile}.

\begin{table}[t]
\centering
\caption{Benchmarks were chosen to represent a mix of batch- and streaming-style computation (marked $\star$).}
\label{tab:eval_benchmarks}
\begin{tabular}{rl}
\toprule
\small\textbf{Name} & \small\textbf{Description}          \\ \midrule
adpcm         & Pulse-code modulation encoder/decoder     \\
bitcoin       & Bitcoin mining accelerator                \\
df            & Double-precision arithmetic circuits      \\
mips32        & Bubble-sort on a 32-bit \mips{} processor \\
nw$^\star$    & DNA sequence alignment                    \\
regex$^\star$ & Streaming regular expression matcher      \\ \bottomrule
\end{tabular}
\end{table}

\section{Evaluation}
\label{sec:eval}

We evaluated \system{} using a combination of \altera{} \de{} \soc{}s and \amazon{}
\fone{} cloud instances.  The \de{}s consist of a Cyclone V
device~\cite{altera2013device} with an 800 MHz dual core \arm{} processor, 
reprogrammable fabric of 110K \lut{}s,  50 MHz clock, and 1 GB of shared DDR3
memory.  \system{}'s \de{} backend was configured to generate bitstreams using
\intel{}'s \quartus{} Lite Compiler and to interact with the \de{}s'
\fpga{} fabric via a soft-IP implementation of an Avalon Memory-Mapped
master.  The \fone{} cloud instances~\cite{F1} support multiple \xilinx{}
UltraScale+ VU9Ps running at 250 MHz and four 16 GB DDR4 channels.  \system{}'s
\fone{} backend was configured to use build tools adapted from the \fone{}
toolchain and to communicate with the instances' \fpga{} fabric over
PCIe.

Table~\ref{tab:eval_benchmarks} summarizes the benchmarks used in our
evaluation, a combination of batch and streaming 
computations.  The ability to handle file \io{} directly from hardware made the
latter easy to support, as developing these benchmarks amounted to repurposing
testbench code designed for functional debugging.  
Benchmarks were compiled prior to running experiments to prime \system{}'s
bitstream caches.  This was appropriate as \system{}'s goal is to provide
virtualization support for applications which have spent sufficient
time in the compile-test-debug cycle to converge on a stable design.

We find that \system{} improves upon \cascade{}'s performance.
Despite targeting a $5\times$ higher frequency on \fone{}, implementing a more
complex program transformation, and accounting for device frequency overheads,
it still achieves a virtual clock frequency~\cite{cascade} within
$\slowdown{}\times$ of native unvirtualized performance and maintains
a reasonable fabric cost.
While non-negligible, these figures do not represent a lower-bound, and
we expect further engineering to reduce them considerably. 

\subsection{Workload Migration}
\label{sec:eval_sr}

\begin{figure}[t]
\centering
\includegraphics[scale=1.0]{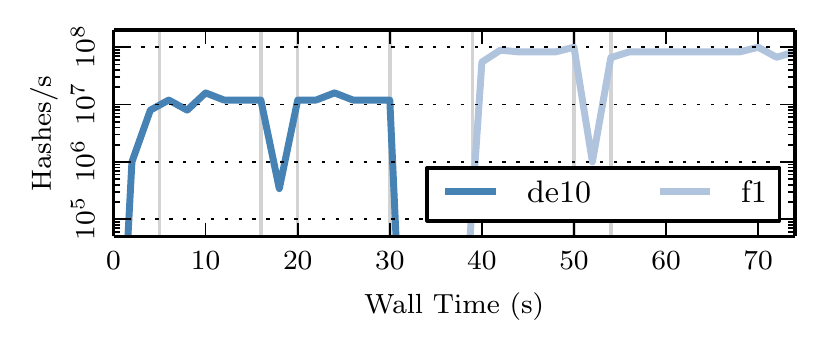}
\caption{Suspend and Resume. \texttt{Bitcoin} is executed on a \de{} target, suspended, saved, and resumed on \fone{}.}
\label{fig:eval_sr}
\end{figure}

Figure~\ref{fig:eval_sr} plots the performance of \texttt{bitcoin} as it is moved
back and forth between software and hardware on two different target
architectures.  This workflow is typical of suspend and resume style
virtualization. The application combines a block of data with a nonce, applies
several rounds of SHA-256 hashing, and loops until it discovers a nonce that
produces a hash under a target value. 

The application begins execution in a new instance of \system{} and, after
running briefly in software, transitions to hardware execution on a \de{}
($t=5$) where it achieves a peak throughput of $16$M nonces evaluated per second.
At ($t=15$) we emit a signal which causes the instance to evaluate a {\tt
\$save} task. Control then transitions temporarily to software as the runtime
evacuates the program's state. The application's
throughput drops significantly during this window, but quickly returns to
steady-state as control returns to hardware ($t=22$). 
\system{} is then terminated ($t=30$), and similar process is initiated on an
\fone{} instance ($t=39$).  In this case, the 
instance evaluates a {\tt restart} task to restore the context which was
saved on the \de{} ($t=50$).  Due to the larger, higher performance hardware on 
\fone{}, the program achieves a higher throughput ($83$M), but suffers from higher
performance degradation during the {\tt restart} as it takes longer to reconfigure.

\begin{figure}[t]
\centering
\includegraphics[scale=1.0]{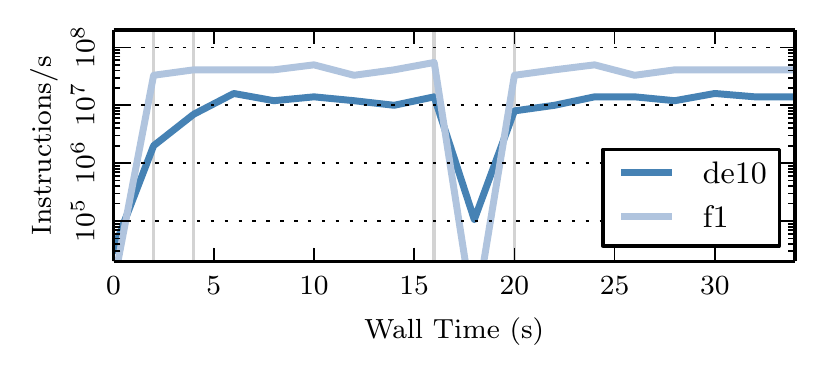}
\caption{Hardware Migration. \texttt{Mips32} begins execution on one target and is migrated mid-execution to another.}
\label{fig:eval_hm}
\end{figure}

Figure~\ref{fig:eval_hm} plots the performance of a single-cycle 32-bit \mips{}
processor consisting of registers, a datapath, and on-chip memory.
The CPU repeatedly randomizes and sorts an in-memory array,
with execution transitioning between two \fpga{}s. The
workload is typical of long-running batch computations which are coalesced to
improve data center utilization.

The curves show two different execution contexts: one where the program is
migrated between nodes in a cluster of \de{}s, and one where it is
migrated between \fone{} instances. The timing of key events is synchronized
to highlight the differences between the environments. In both cases
control begins in software and transitions shortly thereafter to hardware
($t=2,4$) where the targets achieve throughputs of 14M and 41M
instructions per second, respectively. At ($t=15$) we emit a signal which
causes both contexts to evaluate {\tt \$save/\$restart} tasks as the program is
moved between \fpga{}s. A short time later ($t=20$), performance returns to
peak.
Compared to the previous example and the same experiment run on some of the other
benchmarks, the performance degradation during hardware/software transitions is
more pronounced for {\tt mips32}, with the virtual frequency temporarily
lowering to $2$K on \fone{}. This is partially due to the large amount
of state which must be managed by {\tt get/set} requests compared to other
benchmarks (the state of a \mips{} processor consists of its registers,
data memory, and instruction memory).

\begin{figure}[t]
\centering
\includegraphics[scale=1.0]{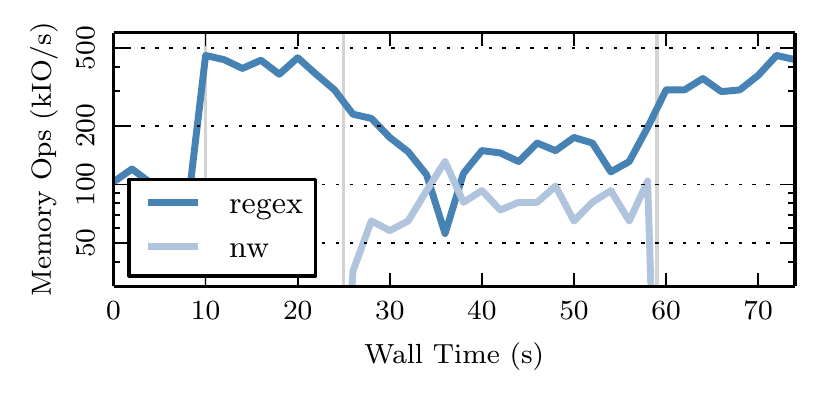}
\caption{Temporal Multiplexing. \texttt{Regex} and \texttt{nw} are time-slice scheduled to resolve contention on off-device \io{}.}
\label{fig:eval_tm}
\end{figure}

\subsection{Multitenancy}

Figure~\ref{fig:eval_tm} plots the performance of two streaming-style
computations on a \de{}. Both read inputs from data files that are too large
to store on-chip. The first (\texttt{regex}) reads in characters and
generates statistics on the stream using a regular expression matching algorithm.
The second (\texttt{nw}) reads in DNA sequences and evaluates how well they match
using a tile-based alignment algorithm.

The regular expression matcher begins execution in a new instance of \system{}
and, at time ($t=10$), transitions to hardware where it achieves a peak
throughput of $500,000$ reads per second. At ($t=15$), the sequence aligner
begins execution in a second instance of \system{}. For the next few seconds,
the performance of the matcher is unaffected. At ($t=24$), the aligner
transitions to hardware and the hypervisor is forced to temporally multiplex the execution
of both applications, as they now contend on a common \io{} path between software
and hardware. During the period where both applications are active ($t=24-60$),
the matcher's throughput drops to slightly less than $50\%$. This is due to the
hypervisor's use of round-robin scheduling and the fact that the primitive read
operations performed by the matcher (characters) require less time to run to
completion than the primitive read operations performed by the aligner
(strings). 

At ($t=60$), the sequence aligner completes execution, and the throughput for the
matcher returns to its peak value shortly thereafter.  Compared to previous
examples, the time required to transition between performance regimes is
slightly more pronounced. This is due to \system{}'s use of adaptive
refinement~\cite{cascade} to determine the time spent in hardware execution
before yielding control back to the \repl{}. It takes several seconds after the
aligner finishes execution for \cascade{} to adjust back to a schedule which
achieves peak throughput while also maintaining interactivity.

\begin{figure}[t]
\centering
\includegraphics[scale=1.0]{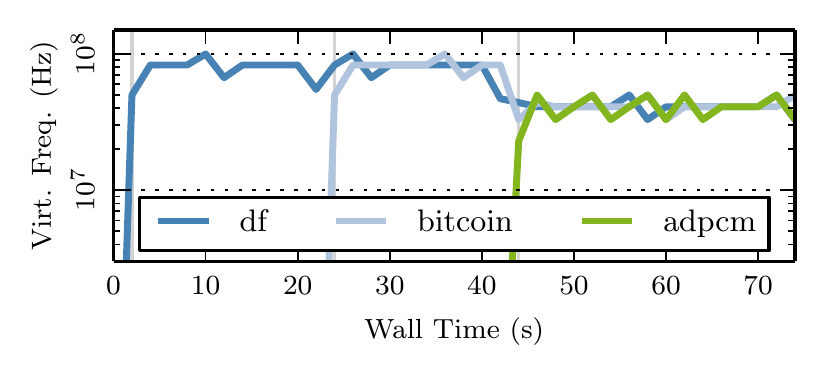}
\caption{Spatial Multiplexing. \texttt{Bitcoin}, \texttt{df}, and \texttt{adpcm}
         are co-scheduled on one device without contention.}
\label{fig:eval_sm}
\end{figure}

Figure~\ref{fig:eval_sm} plots the performance of some batch-style computations
on an \fone{} instance. The first two applications read small inputs sets and transition
to long-running computation before returning a result. The former (\texttt{df}) performs
double-precision floating-point computations characteristic of numeric simulations, and the
latter (\texttt{bitcoin}) is the miner described in~\S\ref{sec:eval_sr}.  Compared to
the previous examples, the results are unremarkable.  
Without resource contention, the hypervisor is able to run both in parallel.
The applications begin software execution in separate instances of \system{}
($t=0,22$) and after transitioning to hardware ($t=2,24$) achieve a virtual
clock rate~\cite{cascade} of $83$ MHz.
At ($t=42$), another batch-style application that encodes and decodes audio
data (\texttt{adpcm}) begins execution in a new instance of \system{}. While
the hypervisor can run this application in parallel with the first two,
lowering its application logic onto the \fone{} instance causes the resulting
design to no longer meet timing at the peak frequency of $250$ MHz.  To
accommodate all three applications, the global clock is set to $125$ MHz,
reducing their virtual clock frequencies to $41$ MHz. The
\system{} hypervisor hides the number of applications running simultaneously
from the user.  As a result, this can lead to unexpected performance
regressions in our prototype. Future work can address this by running each
application in an appropriate clock domain, with clock-crossing logic added
automatically as needed.

\subsection{Quiescence}
\label{sec:qui}

Saving and restoring large volumes of state not only degrades reconfiguration
performance (Figure~\ref{fig:eval_hm}) but also requires a large amount of
device-side resources to implement (\S~\ref{sec:over}).
\system{}'s quiescence interface allows developers to signal when a program is
quiescent and which variables are stateful at that time.
We found that most of our benchmarks had a large number of volatile variables,
including $99\%$, $96\%$, and $71\%$ of \texttt{df}'s, \texttt{bitcoin}'s, and \texttt{mips32}'s state.
For these applications, implementing quiescence resulted in an average LUT and FF
savings of $50\%$ and $15\%$, respectively.
In our other benchmarks, $1/8$ to $1/4$ of the state was volatile.
Implementing quiescence for them resulted in an average LUT and FF savings of
$2\%$ and $9\%$, respectively.

\subsection{Overheads}
\label{sec:over}

There are two major sources of overheads in programs constructed by \system{}.
The first are discrete, non-fundamental overheads resulting from how programs
are virtualized in hardware in the \system{} prototype.  Implementing the
semantics of the original program with the ability to pause execution in the
middle of a virtual clock cycle involves toggling the virtual clock variable,
evaluating relevant program logic, and latching variable assignments.
When these are done in separate hardware cycles, there is a minimum
$3\times$ performance overhead.  This is an artifact of our implementation
rather than a fundamental requirement and can be improved with future work on
target-specific backends.

The second source of overheads comes from the state access and execution
control logic added by \system{}, which has a less-obvious impact on designs
targeting \fpga{} hardware.  To evaluate these overheads, we compile our
benchmarks under a number of different conditions and measure the resource
usage and achieved device frequency. We perform these compilations using Vivado
on \fone{}, target \amorphos{}'s maximum frequency (250 MHz), and use the
reported delay to determine the frequency achieved.

As a baseline, we compile our benchmarks natively on \amorphos{}, providing an
upper bound on resource and frequency overheads.  We also simulate a Cascade on
\amorphos{} baseline by compiling our benchmarks without system tasks, which
avoids overheads introduced by our new state machine transformations.  Finally,
we modified our benchmarks to implement the quiescence protocol, allowing us to
estimate the savings of exposing reconfiguration to developers and establishing
a lower bound on state access overhead.  Due to the complexity of fully
rewriting our benchmarks for these cases, we only focus on replicating
overheads and not functionality.

\begin{figure}[t]
\centering
\includegraphics[width=\columnwidth]{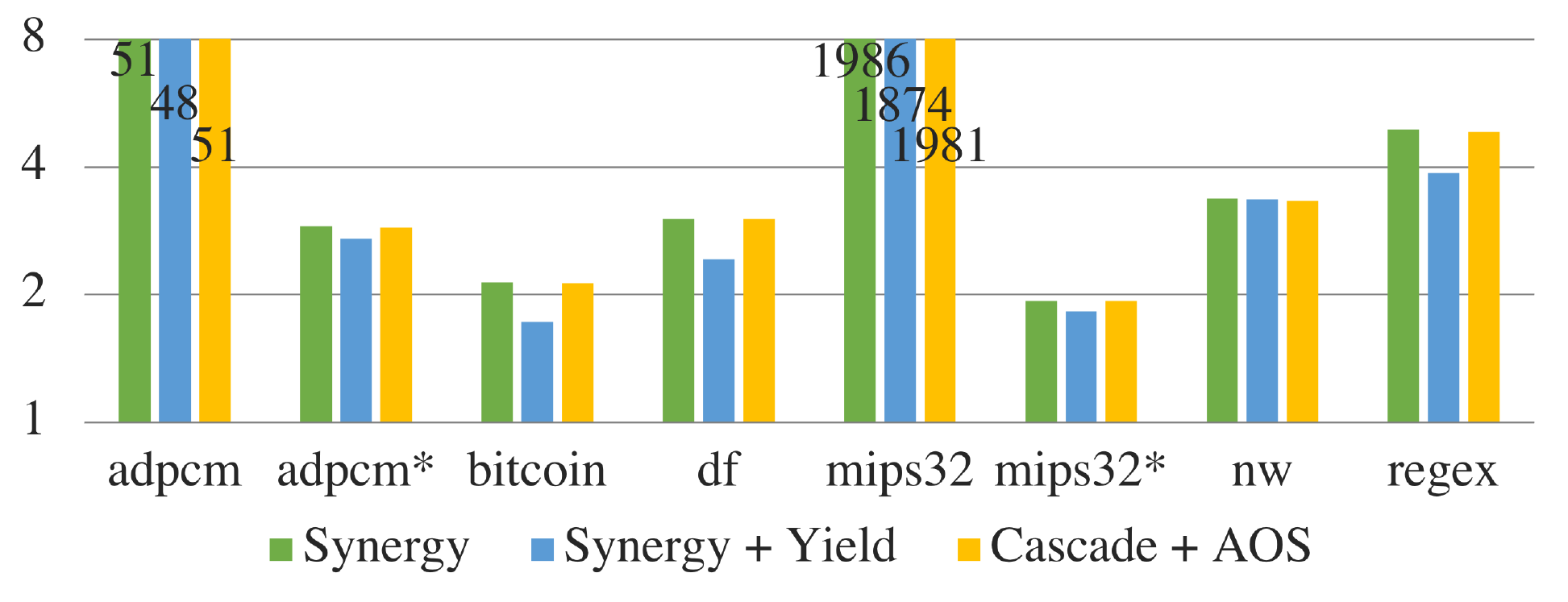}
\caption{FF usage normalized to that of \amorphos{}.}
\label{fig:eval_ff}
\end{figure}

\begin{figure}[t]
\centering
\includegraphics[width=\columnwidth]{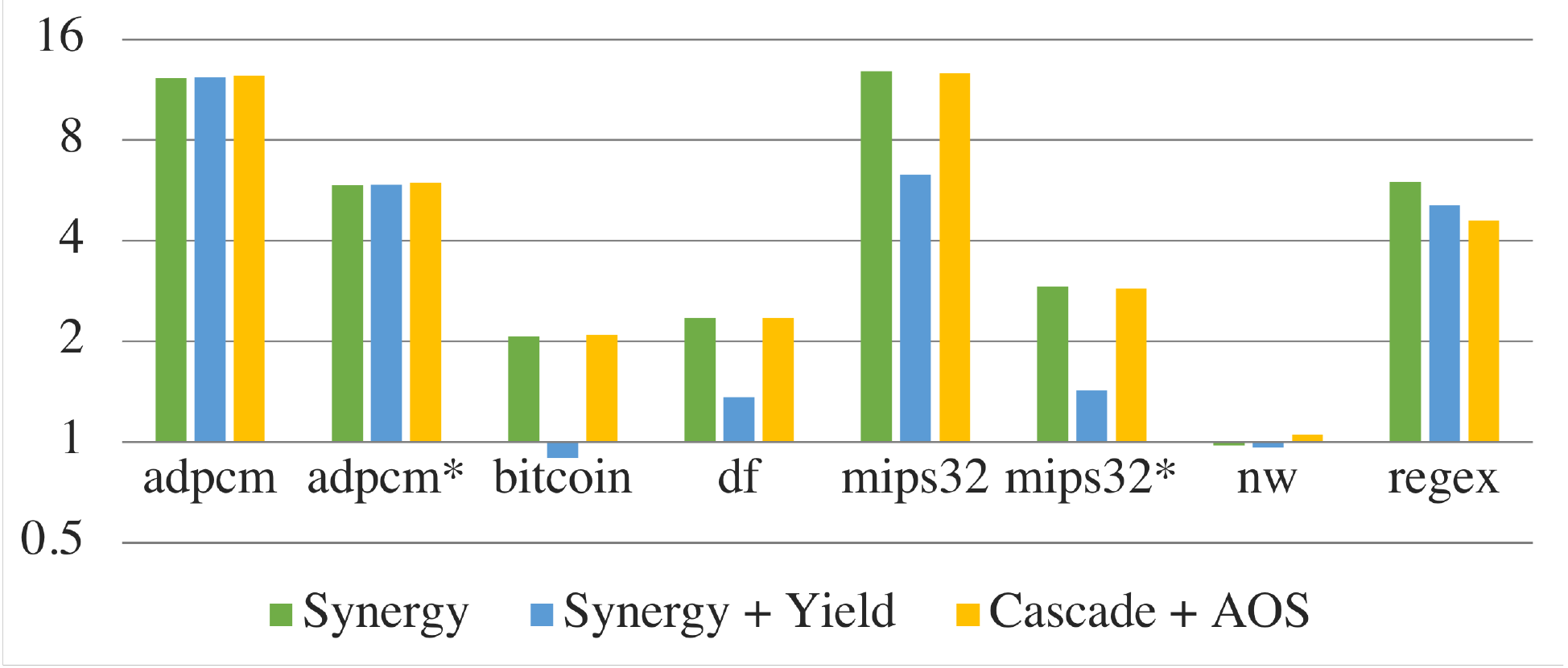}
\caption{LUT usage normalized to that of \amorphos{}.}
\label{fig:eval_lut}
\end{figure}

Figures~\ref{fig:eval_ff} and \ref{fig:eval_lut} show \system{}'s FF and LUT
usage is generally $2-4\times$ and $1-6\times$ native, respectively.
For {\tt adpcm} and {\tt mips32}, the results exceed the height of the graph
and have been labeled with the appropriate values.
These two benchmarks are exceptional due to their use of large
on-chip RAMs, which Vivado implements using FFs under \system{} instead of LUTRAMs.
Creating RAMs out of FFs can also require additional LUTs to implement muxing logic.
The {\tt adpcm}* and {\tt mips32}* results compare against \amorphos{} using FFs for RAMs and show
that \system{}'s overheads are resonable under these conditions.
Future work should enable state access transformations that preserve the
memories in user designs, which would eliminate this problem.
Overall, we find that \system{}'s overheads are similar to
Cascade's and that using quiescence annotations can provide savings
of up to $\sim$$2\times$.

\begin{figure}[t]
\centering
\includegraphics[width=\columnwidth]{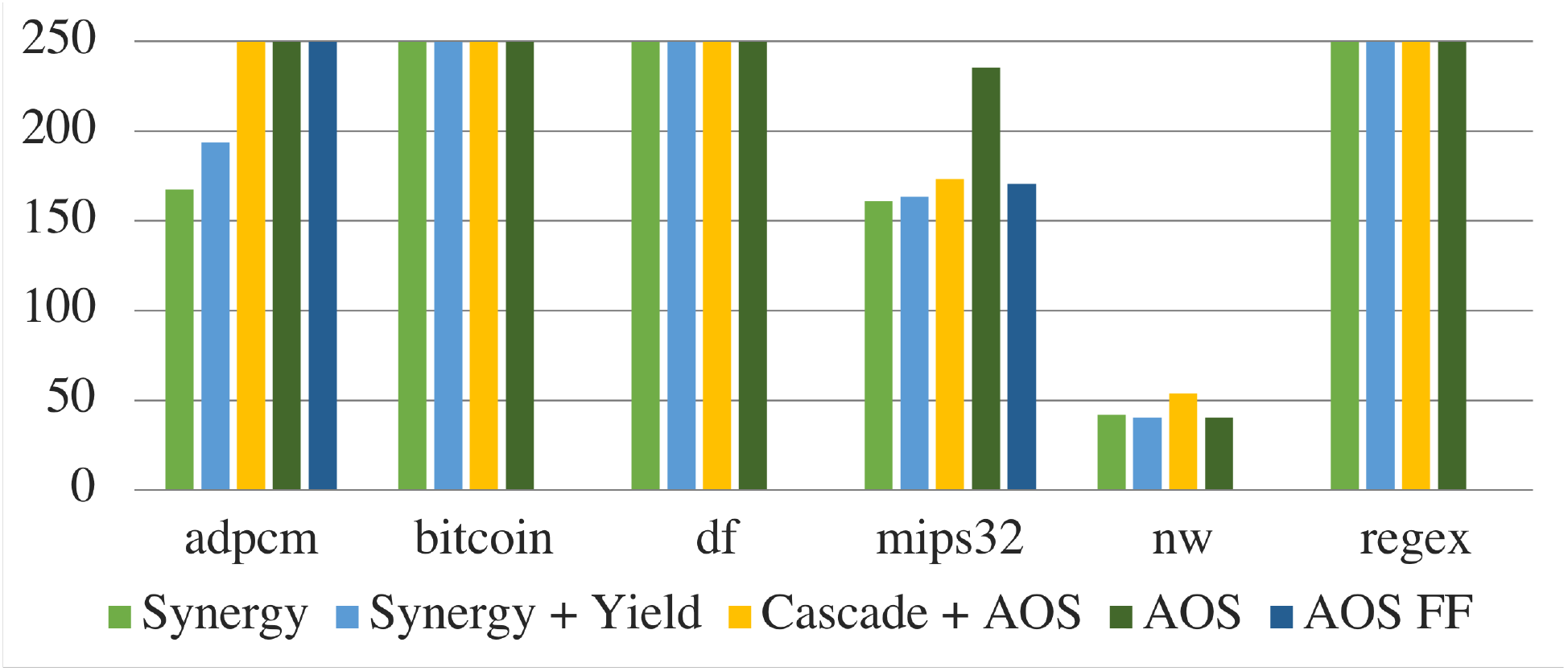}
\caption{Design frequency achieved in MHz.}
\label{fig:eval_freq}
\end{figure}

Figure~\ref{fig:eval_freq} shows that \system{} does not reduce the design's
operating frequency in most cases.
However, {\tt adpcm} is an exception, likely due to its use of system tasks from inside its complex
control logic, which makes execution control much more expensive to implement.
We see that \system{}'s frequency overhead for {\tt mips32} is almost
entirely due to forcing the use of FFs to implement RAMs.
When normalized against \amorphos{} using FFs (AOS FF), \system{} was less
than $6\%$ slower despite supporting full state capture.
We find that for {\tt nw}, \system{} and the Cascade baseline
achieve higher frequencies than native, likely due to {\tt nw}'s complexity
creating a higher-than-normal volatility in compiler outcomes.
When combined with the previous $3\times$ overhead, we find that \system{}'s
overall execution overhead is within $\slowdown\times$ that of native.

While not shown in the data above, we tried compiling {\tt adpcm} and {\tt nw} with an
anti-congestion strategy to see if it would help with their complex designs.
We found that this improved both their frequencies under \system{} by $47\%$.
With quiescence annotations, {\tt adpcm} still improved $23\%$ and {\tt nw}, $37\%$.
Applying the same strategy to {\tt nw} under AOS only gave an improvement of $26\%$.
This indicates that optimizing compiler strategies could be a great avenue
for offsetting the costs of code transformations.

\section{Limitations}
\label{sec:limit}

\paragraph{Source Code}
\system{} guarantees isolation between runtime instances with respect to
hardware execution.  However information leak between distrustful parties through 
source code and side channels is still possible.  Leaks through sharing source
could be mitigated with cryptographically secure channels,
but \system{} would still function as a trusted party.

\paragraph{Compilation Cache}
\system{}'s backends rely on compilation caches to
reduce overhead in production environments, 
as the alternative would be to transition back to software 
and wait through recompilation on virtualization events.
\system{} uses techniques such as deterministic code generation to increase cache hit rates.
As more applications use \fpga{}s, cache hit rates may drop 
and symmetry-breaking or speculative compilation may be needed to 
compensate.

\section{Related Work}
\label{sec:rel}

\paragraph{\fpga{} \os{} and Virtualization}
Primitives for \fpga{}s include sharing \fpga{} fabric
~\cite{chen14cloud,byma14openstack,fahmy15cloudcom,fpgadc15,knodel15,knodel16},
spatial multiplexing
~\cite{fu-scheduling-08,os-fpga-rt-04,wassi-multi-shape-tasks-14,
chen12scheduling}, context switch~\cite{Lee2010HardwareCM,rupnow09block},
memory virtualization~\cite{coram,leap,matchup,reconOS},
relocation~\cite{kalte-context-05}, 
preemption~\cite{levinson-preemptive-00}, and interleaved hardware-software
task execution~\cite{xilinx-OS, os-fpga-rt-04,
wassi-multi-shape-tasks-14,gonzalez-12-virt}.  Core techniques include
virtualizing \fpga{} fabric, including regions~\cite{codezero},
tasks~\cite{plessl2005zippy}, processing elements~\cite{score}, IPC-like
communication primitives~\cite{tartan06asplos}, and abstraction
layers/overlays~\cite{virtualRC,huang09fpgavirt,brant2012zuma,rcmw,intermediate-fabrics}

Extending \os{} abstractions to \fpga{}s is an area of active research.
ReconOS~\cite{reconOS} extends eCos~\cite{domahidi2013ecos} 
with \emph{hardware threads} similar to Hthreads~\cite{hthreads}.
Borph~\cite{borph, borph-08} proposes a \emph{hardware process} abstraction. 
Previous multi-application \fpga{} sharing proposals~\cite{chen12scheduling,remap,fuse,werner12virt}
restrict programming models or fail to provide isolation.
\os{} primitives have been combined to form OSes for
\fpga{}s~\cite{reconOS,borph,borph-08,murac} as well as \fpga{}
hypervisors~\cite{codezero,plessl2005zippy,tartan06asplos,score}.  Chen et al.
explore virtualization challenges when \fpga{}s are a shared
resource~\cite{chen14cloud}; \amorphos{}~\cite{amorphos} provides an
\os{}-level management layer to concurrently share \fpga{}s among mutually
distrustful processes.  
ViTAL~\cite{vital} exposes a single-\fpga{} abstraction for scale-out acceleration 
over multiple FPGAs; unlike \system{}, it exposes a homogeneous 
abstraction of the hardware to enable offline compilation.
The Optimus~\cite{optimus} hypervisor supports
spatial and temporal sharing of FPGAs attached to the host memory
bus, but does virtualize reconfiguration capabilities.
Coyote~\cite{coyote20osdi} is a shell for FPGAs 
which supports both spatial and temporal multiplexing as well as 
communication and virtual memory management. While sharing goals 
with these systems, \system{} differs fundamentally from
them by virtualizing FPGAs at the language level \emph{in addition}
to providing access to OS-managed resources.

\paragraph{\fpga{} Programmability and Compilation}
\hdl{}s, primarily Verilog~\cite{verilog} and VHDL~\cite{vhdl}, have served as
the lowest abstraction level and least common denominator for programming
\fpga{}s for decades. Improving \fpga{} programmability through higher level
languages~\cite{pyrtl,opencl,lime,AutoESL,MARC,sdaccel, Anderson06enablinga,carte,
lebak2005parallel,spatial}, domain-specific
languages~\cite{linqits,MARC,fpmr,chisel,vforce,olaf-2016,autoaccel16isca,plasticine},
or frameworks~\cite{opencl,lime,AutoESL} is an active area of research, which
while orthogonal, can greatly benefit from our contributions.
AccelNet~\cite{accelnet} supports fast
in-network packet processing on top of Microsoft Catapult SmartNICs.
FlexNIC~\cite{flexnic, flexnic:hotos} is a programmable networking device architecture to offload 
packet processing tasks. NICA~\cite{nica} uses FPGA-based SmartNICs to accelerate network servers. 
Floem~\cite{floem} is compiler to simplify offload development on SmartNICs.
 
\fpga{} compilers incur significant overhead. 
\system{} adapts many overhead-reduction techniques from the software domain such as
eliminating redundant re-compilation~\cite{ccache,distcc,icecream}, distributed build
caching~\cite{cloudbuild,bazel,vesta}, and \jit{} compilation~\cite{nodejs}.
\fpga{} compilation can be further improved with a
virtualization layer.  Overlay-based virtualization ~\cite{brant2012zuma,
reconos-zuma, hoplite, overlay-isa, virtualRC,rcmw} abstracts away
target-specific details and enables fast compilation and lower deployment
latency. The approach reduces utilization and performance. \system{} 
and ~\cite{cascade} work at this level, while
AmorphOS~\cite{amorphos} works at the application-\os{} boundary.

\paragraph{Hardware-Software Partitioning}
The \fpga{} design cycle relies on developing a design in a
high-fidelity simulator~\cite{quartus,vivado} before compiling to hardware.
Simulation incurs order of magnitude slowdowns compared to hardware execution,
but is necessary for debugging. Many attempts have been made to bridge this
performance gap, including systems which enable migration between different
speed simulators~\cite{palladium}, higher-level languages with partitioned
runtime environments~\cite{lime}, and \os{}-managed communication
channels~\cite{leap}. \system{} builds on the approach taken by \cascade{} as it
does not require explicit interface changes~\cite{lime,leap} or hardware
support~\cite{palladium}.

\section{Conclusion}
\label{sec:conc}

\fpga{}s are emerging in data centers so
techniques for virtualizing them are urgently needed to enable
them as a practical resource for on-demand hardware acceleration. \system{}
is a compiler/runtime solution that supports
multi-tenancy and workload migration on hardware which is available \emph{today}.

\begin{acks}
We thank the PC and our shepherd Nate Foster for their insightful feedback.
This research was supported by NSF grants CNS-1846169 and CNS-2006943, and U.S.
Department of Energy, National Nuclear Security Administration Award Number DE-NA0003969.
\end{acks}

\appendix
\section{Artifact Appendix}
\balance

\subsection{Abstract}

This artifact appendix documents the requirements and instructions for setting up \system{}
and how to reproduce the results of the experiments presented in our ASPLOS'21 paper.
Code for our various backends can be found at:
\begin{itemize}
\item {AWS F1}: \url{https://github.com/JoshuaLandgraf/cascade}
\item {SW, DE10-Nano}: \url{https://github.com/eschkufz/cascade}
\end{itemize}

\subsection{Artifact Checklist}

Checklist details: \url{https://ctuning.org/ae/checklist.html}
{\small
\begin{itemize}
  \item {\bf Algorithm}: Hardware-accelerated virtualized Verilog runtime
  \item {\bf Program}: Assorted Verilog programs provided in our repos
  \item {\bf Compilation}: GCC on Linux, Clang on macOS, AWS FPGA Dev AMI (includes Vivado) for AWS F1, Quartus Lite for DE10-Nano backend
  \item {\bf Binary}: Our software is compiled from source. Vivado and Quartus binaries are provided through Amazon and Intel, respectively.
  \item {\bf Data set}: Example data files provided with benchmarks
  \item {\bf Run-time environment}: AWS F1 backend runs on AWS FPGA Dev AMI 1.7. DE10-Nano backend runs on Ubuntu 20 (VMs supported). SW backend can run on Ubuntu 20 or macOS 10.15.
  \item {\bf Hardware}: AWS F1 instance or DE10-Nano kit.
  \item {\bf Run-time state}: FPGA bitstreams cached for use in repeat execution.
  \item {\bf Execution}: Benchmarks can run for minutes on hardware backends. Initial Quartus builds take $\sim$20 minutes each; Vivado builds take $\sim$2 hours, but large, timing-contrained builds can take several times that.
  \item {\bf Metrics}: \system{} can profile virtual application frequency.
  \item {\bf Output}: Profiling data is output to the console or a log file.
  \item {\bf Experiments}: We provide a guide for environment setup, software installation, and experimental methodology for F1 instances.
  \item {\bf How much disk space required?}: Our own software uses $\sim$100MB. AWS FPGA Dev AMI uses 75GB with builds using $\sim$1GB each. Quartus Lite uses 15GB with builds using 100s of MB each.
  \item {\bf How much time is needed to prepare workflow?}: Software can be set up on Ubuntu or macOS in $\sim$10 minutes. Setting up the AWS F1 and DE10-Nano environments can take 1-3 hours.
  \item {\bf How much time is needed to complete experiments?}: Our F1 experiments can take 2 days, mostly for performing FPGA builds. Our DE10 experiments could take 2 hours.
  \item {\bf Publicly available?}: Yes.
  \item {\bf Code licenses?}: BSD-2.
  \item {\bf Workflow framework used?}: Scripting via our software's library interface is supported.
\end{itemize}

\subsection{Description}

\subsubsection{How to Access}

Our code can be obtained by cloning the repositories linked in the abstract.
The AWS F1 repo contains a preview of our F1 backend, which currently requires a manual setup process.
The main repo is recommended for evaluating all other backends.

\subsubsection{Hardware Dependencies}

The AWS F1 backend requires an AWS f1.2xlarge instance, or f1.4xlarge for FPGA migration experiments.
The DE10-Nano backend requires a DE10-Nano kit, available at \url{http://de10-nano.terasic.com}.

\subsubsection{Software Dependencies}

The AWS F1 backend requires the AWS FPGA Developer AMI, which includes all proprietary software needed.
The DE10-Nano backend requires Intel's Quartus Lite software, which may require making a free account with Intel.
All other dependencies are open-source and covered in our installation guides.

\subsection{Installation}

A thorough setup guide is available for the F1 backend:
\url{https://github.com/JoshuaLandgraf/cascade/blob/artifact/ARTIFACT.md}.
Otherwise, the READ\-ME covers installation on Ubuntu and macOS, obtaining Intel's Quartus software, and setting up a DE10-Nano:
\url{https://github.com/eschkufz/cascade/blob/master/README.md}.

\subsection{Experiment Workflow}

Our primary experiments consist of running several benchmarks on \system{} and
demonstrating the ability to save/restore state, migrate execution across FPGAs,
and share hardware resources.
This can be accomplished from the command line, with directives specified dynamically through the REPL,
or through scripting via the C++ library interface.
\system{} tracks when these directives are executed as well as the virtual application's frequency
and can print this information to the console or log it to a file.

\subsection{Evaluation and Expected Results}

In order to help simplify the process of running experiments,
we provide several C++ programs that use \system{}'s library interface to automate experiment execution.
These can be found in the artifact branch of the AWS F1 repo, in the experiments folder.
The process of building and running these experiments is documented at
\url{https://github.com/JoshuaLandgraf/cascade/blob/artifact/experiments/README.md}.
We currently provide two versions of the experiments, one for native F1 execution,
and one for software simulation.
The expected results are displayed in this paper and can be compared with the
performance data obtained by running the experiments.

\subsection{Experiment Customization}

Since \system{} is a runtime, it enables a wide variety of experiments beyond those shown in the paper.
It is especially simple to look over the experiments provided and tweak them to run for longer,
use different or more benchmarks, or execute more complex sequences of operations.
Many of our benchmarks already have different top-level files that tweak the inputs,
actions performed, or number of execution iterations.
Users can also modify the provided benchmarks or run their own Verilog programs on \system{}.

\bibliographystyle{ACM-Reference-Format}
\bibliography{paper}

\end{document}